\documentclass[aps,prd,twocolumn,10pt,superscriptaddress,nofootinbib,nobibnotes,longbibliography]{revtex4-1}

\usepackage{amssymb}
\usepackage{graphicx}
\usepackage{amsmath}
\usepackage{hyperref}
\usepackage{subfigure}
\usepackage{multirow}
\usepackage{setspace}
\usepackage{verbatim}
\usepackage{float}
\usepackage{color}
\usepackage{ulem}
\usepackage[utf8]{inputenc}
\usepackage[table,xcdraw]{xcolor}
\usepackage{makecell}
\usepackage{url}
\usepackage{bm}

\begin{document}

\title{Broadband Chiral Primordial Gravitational Waves from Constant-roll Inflation in Parity-violating Symmetric Teleparallel Gravity}

\author{Xu Zhang}
\affiliation{Department of Physics, Anhui Normal University, Wuhu, 241002, China}
\affiliation{Center for Astrophysics and Astronomical Technology, Anhui Normal University, Wuhu, 241002, China}

\author{Chang Liu} 
\email[]{liuchang@yzu.edu.cn}
\affiliation{Center for Gravitation and Cosmology, College of Physical Science and Technology, Yangzhou University, Yangzhou, 225009, China}

\author{Chengjie Fu}
\email[]{fucj@ahnu.edu.cn}
\affiliation{Department of Physics, Anhui Normal University, Wuhu, 241002, China}
\affiliation{Center for Astrophysics and Astronomical Technology, Anhui Normal University, Wuhu, 241002, China}

\begin{abstract}
We investigate primordial gravitational waves (GWs) generated during constant-roll inflation in a parity-violating extension of symmetric teleparallel gravity. The parity-violating interactions leave the background evolution and linear scalar perturbations unchanged, while inducing velocity birefringence in the tensor sector. Consequently, one of the two circular polarization states undergoes tachyonic amplification, producing a strongly blue and nearly fully chiral tensor spectrum from the cosmic microwave background (CMB) to interferometer scales. We identify viable constant-roll parameter regions consistent with current CMB constraints and determine the largest coupling strength compatible with both CMB B-mode measurements and the LIGO-Virgo-KAGRA (LVK) O1--O4a bound on the stochastic GW background. The predicted CMB B-mode spectra may be detectable by LiteBIRD, while the enhanced high-frequency signal could be accessible to the LISA--Taiji network and LVK O5 run. The model also predicts nonvanishing TB and EB correlations. These results highlight the potential of combining CMB and multi-band GW observations to test parity-violating gravity during inflation.
\end{abstract}

\maketitle

\section{Introduction}
Cosmic inflation provides a simple framework for resolving the horizon and flatness problems while generating the primordial fluctuations that seed the large-scale structure of the Universe~\cite{Starobinsky:1980te,Guth:1980zm,Linde:1981mu,Albrecht:1982wi,Linde:1983gd,Guth:1982ec,Starobinsky:1982ee,Bardeen:1983qw,Kodama:1984ziu}. Among its key predictions, primordial gravitational waves (GWs) are particularly valuable, as their amplitude and scale dependence offer probes of the inflationary energy scale and the underlying theory of gravity~\cite{Guzzetti:2016mkm,Caprini:2018mtu,Campeti:2020xwn}. Searches for their imprint on
the cosmic microwave background (CMB) have placed stringent limits on the amplitude of primordial tensor perturbations on large scales~\cite{BICEP:2021xfz,AtacamaCosmologyTelescope:2025nti}. At much smaller scales, pulsar timing arrays and ground- and space-based interferometers are opening complementary frequency windows across a broad range of frequencies. However, the nearly scale-invariant tensor spectrum predicted by conventional single-field slow-roll inflation is generally too weak to be detected by interferometers. This has motivated the study of inflationary mechanisms capable of producing a blue-tilted tensor spectrum or a localized enhancement on small scales~\cite{Guzzetti:2016mkm,Cook:2011hg,Mukohyama:2014gba,Cai:2016ldn,Bartolo:2016ami,Mylova:2018yap,Satoh:2007gn,Fu:2020tlw,Cai:2021uup,Fu:2023aab,Zhai:2025flf}.

Parity violation in gravity provides a particularly distinctive realization of this possibility. When the two circular polarization states of GWs obey helicity-dependent equations of motion, velocity or amplitude birefringence can generate a chiral primordial GW background~\cite{Lue:1998mq,Alexander:2009tp,Wang:2012fi,Alexander:2016hxk,Bartolo:2017szm,Qiao:2019hkz,Zhu:2022dfq}. Such a background can be tested not only through its total GW intensity, but also through the CMB cross-correlation spectra TB and EB~\cite{Saito:2007kt,Gluscevic:2010vv}, as well as through potential measurements of circular polarization by networks of space-based interferometers~\cite{Seto:2020zxw,Orlando:2020oko,Chen:2024ikn}. 
Combining these observables across widely separated scales therefore provides a powerful way of distinguishing parity-violating physics from parity-symmetric sources of a stochastic GW background.

An economical framework for parity violation is symmetric teleparallel gravity (STG), in which gravity is encoded in the nonmetricity of a flat and torsion-free connection. Within this framework, the Symmetric Teleparallel Equivalent of General Relativity (STEGR) is dynamically equivalent to general relativity (GR) and yields the same field equations~\cite{Nester:1998mp,BeltranJimenez:2019esp,Capozziello:2022zzh}. Parity-violating extensions of STEGR can be constructed by coupling the scalar field to parity-odd terms that are quadratic in the nonmetricity tensor~\cite{Li:2021mdp,Li:2022vtn}. On a spatially flat Friedmann-Robertson-Walker (FRW) background, these operators leave both the background evolution and linear scalar perturbations unchanged, while modifying the vector and tensor sectors. In particular, they induce velocity birefringence in tensor perturbations. A recent application of this mechanism to axion inflation demonstrated that the rapid evolution of the coupled scalar field, which also drives inflation, through a steep feature in its potential can generate a localized, multipeaked, and highly chiral GW signal within the frequency band of space-based interferometers~\cite{Zhai:2025flf}. In that realization, the parity-violating correction was chosen to be negligible on CMB scales. It is therefore natural to ask whether a different inflationary background can sustain parity-violating effects in primordial GWs over a much broader range of wavelengths, thereby making both CMB and interferometer observations simultaneously relevant.

Constant-roll inflation provides such a background. It relaxes the usual slow-roll assumption by requiring the inflaton to maintain a constant rate of roll~\cite{Motohashi:2014ppa,Motohashi:2017aob,Cicciarella:2017nls,Yi:2017mxs,Gao:2018cpp,GalvezGhersi:2018haa,Gao:2019sbz,Lin:2019fcz,Motohashi:2019rhu}. This framework generalizes slow-roll inflation and admits exact solutions that are compatible with current observational constraints~\cite{Motohashi:2025qgd}. In this work, we combine constant-roll inflation with a parity-violating extension of STEGR and study its tensor phenomenology from CMB to interferometer scales.
 
The remainder of this paper is organized as follows. In Sec.~II, we review the constant-roll inflation and discuss its current observational constraints. In Sec.~III, we introduce the parity-violating STG framework, derive the tensor mode equations, and present the resulting CMB and present-day GW spectra. Section~IV is devoted to our conclusions. Throughout this work, we adopt the natural units with $c=\hbar=1$.

\section{Constant-roll inflation}
Constant-roll inflation is a phenomenological class of inflationary models in which the usual slow-roll condition is relaxed and replaced by the requirement that the inflaton’s rate of roll remains constant. In this work, we focus on constant-roll dynamics in single-field canonical inflation, described by the action
\begin{align}
    S=\int {\rm d}^4x\sqrt{-g}\left[ \frac{M^2_{\rm Pl}}{2}R-\frac{1}{2}g^{\mu\nu}\partial_\mu\phi\partial_\nu\phi-V(\phi)\right],
\end{align} 
where $M_{\rm Pl}$ denotes the reduced Planck mass. For a spatially flat FRW background, ${\rm d}s^2=-{\rm d}t^2 + a(t)^2\delta_{ij}{\rm d}x^i{\rm d}x^j$ with $a(t)$ being the scale factor, the background dynamics are governed by
\begin{align}\label{background_equation}
3M_{\rm Pl}^2H^2 = \frac{1}{2}\dot\phi^2+V(\phi),\qquad \ddot\phi+3H\dot\phi+V_\phi=0,
\end{align}
with $V_\phi\equiv {\rm d}V/{\rm d}\phi$.
Constant-roll inflation is then characterized by the condition $\ddot\phi={\beta H\dot\phi}$, where $\beta$ is a 
constant parameter. 

A simple potential that realizes this class of inflationary evolution is given by~\cite{Motohashi:2014ppa}
\begin{align}\label{potenial}
    V(\phi) = 3M^2M_{\rm Pl}^2\left[ 1- \frac{3+\beta}{6}\left\{ 1-\cos\left(\sqrt{2\beta}\frac{\phi}{M_{\rm Pl}}\right)\right\}\right],
\end{align}
where $M$ sets the inflation energy scale and is determined by the amplitude of the primordial scalar spectrum. We restrict our discussion to the branch $\phi\geq 0$. Along this branch, the inflaton asymptotically approaches the origin in the infinite past and subsequently rolls towards larger field values. The potential has a form reminiscent of natural inflation, but is shifted downward by a negative constant contribution. In particular, its minimum lies at a negative value, $V_{\rm min}=-\beta M^2M_{\rm Pl}^2$, so that the potential becomes negative beyond a finite field value. The point at which the potential crosses zero is determined by $V(\phi_{\rm c})=0$, yielding
\begin{align}
    \phi_{\rm c} = M_{\rm Pl}\sqrt{\frac{2}{\beta}}\arcsin\sqrt{\frac{3}{3+\beta}}.
\end{align}
Since the effective single-field description would otherwise evolve into a region with $V<0$, it must be supplemented by a mechanism that terminates constant-roll inflation before this point is reached. We therefore assume that the potential is truncated, or equivalently modified by some phase-transition-like dynamics, at a field value $\phi_0$ satisfying $\phi_0<\phi_{\rm c}$. This cutoff marks the end of the constant-roll stage. Let $\phi_i$ denote the field value at which the CMB pivot scale exits the Hubble horizon.  

It is convenient to introduce the {\it e}-folding number measured relative to the critical field value $\phi_{\rm c}$~\cite{Motohashi:2025qgd},
\begin{align}
    N(\phi) \equiv \ln\frac{a(\phi)}{a(\phi_{\rm c})} = \frac{1}{\beta}\ln\left[ \sqrt{\frac{3+\beta}{3}}\sin\left( \sqrt{\frac{\beta}{2}} \frac{\phi}{M_{\rm Pl}}\right)\right].
\end{align}
On the branch $0<\phi<\phi_{\rm c}$, $N(\phi)$ is negative and increases monotonically towards zero as the inflaton evolves towards larger field values. It should be emphasized that $N(\phi)$ does not directly represent the number of {\it e}-folds remaining until the end of constant-roll stage at $\phi_0$. Instead, it measures the logarithmic expansion relative to the hypothetical time at which the inflaton would reach the critical field value $\phi_{\rm c}$. For a fixed value of $\beta$, the critical field value $\phi_{\rm c}$ is uniquely determined, whereas the cutoff $\phi_0$, at which the constant-roll phase is assumed to terminate, depends on the unspecified exit mechanism and may generally take any value in the range $0<\phi_0<\phi_{\rm c}$. The number of {\it e}-folds between the horizon exit of the CMB pivot scale at $\phi_i$ and the end of constant-roll stage at $\phi_0$ is $\Delta N_{\rm CR}=N(\phi_0)-N(\phi_i)$. If inflation continues after this transition, the total number of $e$-folds before the actual end of inflation is $\Delta N_{\rm CR}+\Delta N_{\rm post}$. Since the post-constant-roll evolution is not specified here, $\Delta N_{\rm CR}$ is not required to lie in the conventional range of $50$--$60$.

\begin{figure}
\includegraphics[width=1\columnwidth]{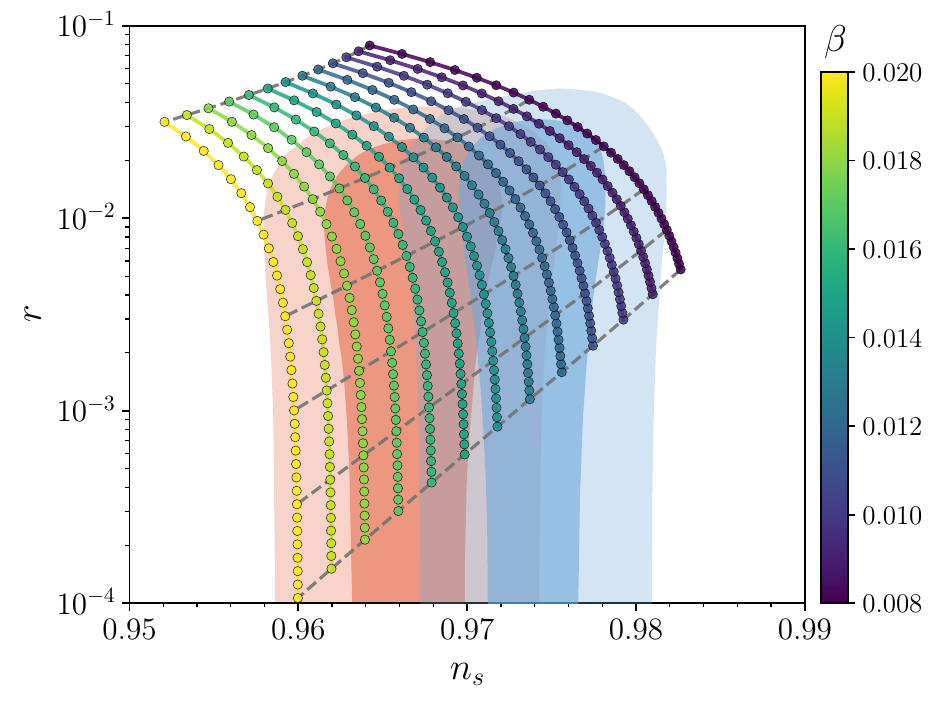}
\caption{ The theoretical predictions in the $n_s-r$ plane at the pivot scale $k_\ast=0.05{\rm Mpc}^{-1}$ for the constant-roll inflation with the potential given in \eqref{potenial}. The points correspond to $\beta$ ranging from $0.008$ to $0.02$ in increments of $0.001$, ordered from right to left, and to $N(\phi_i)$ ranging from $-60$ to $-200$ in increments of $-4$, ordered from top to bottom. The dashed lines connect points with the same value of $N(\phi_i)$.
The red-shaded contours show the observational constraints obtained from the joint analysis of Planck and BICEP/Keck data~\cite{BICEP:2021xfz}. The blue-shaded contours represent the constraints derived from the combined analysis of Planck and ACT data, including CMB lensing and DESI BAO measurements, together with BICEP/Keck data~\cite{AtacamaCosmologyTelescope:2025nti}. }
\label{fig1}
\end{figure}

Within the slow-roll regime, characterized by
\begin{align}
\epsilon_1\equiv-\frac{{\rm d}\ln H}{{\rm d}\ln a}\ll 1, \quad  \epsilon_2 = \frac{{\rm d}\ln \epsilon_1}{{\rm d}\ln a}\ll 1,
\end{align}
the scalar spectral index $n_s$ and the tensor-to-scalar ratio $r$ are given by
\begin{align}
    n_s = 1-2\epsilon_1-\epsilon_2,\qquad r=16\epsilon_1.
\end{align}
For the constant-roll background considered here, the Hubble parameter can be expressed as
\begin{align}
    H=M\sqrt{1-h} , \qquad h=\frac{3e^{2\beta N}}{3+\beta}.
\end{align}
The corresponding Hubble-flow parameters are
\begin{align}
    \epsilon_1 = \frac{\beta h}{1-h},\qquad \epsilon_2 =\frac{2\beta}{1-h}.
\end{align}
It then follows that~\cite{Motohashi:2025qgd}
\begin{align}
    n_s=1-2\beta\frac{1+h}{1-h}, \qquad r=16\beta\frac{h}{1-h},
\end{align}
where the right-hand sides are evaluated when the CMB pivot scale, $k_\ast=0.05{\rm Mpc}^{-1}$, exits the Hubble horizon, corresponding to $N(\phi_i)\equiv N_\ast$. The two inflationary observables are therefore determined by $\beta$ and $N_\ast$, or equivalently by $\beta$ together with the field value $\phi_i$.

\begin{figure}
\includegraphics[width=1\columnwidth]{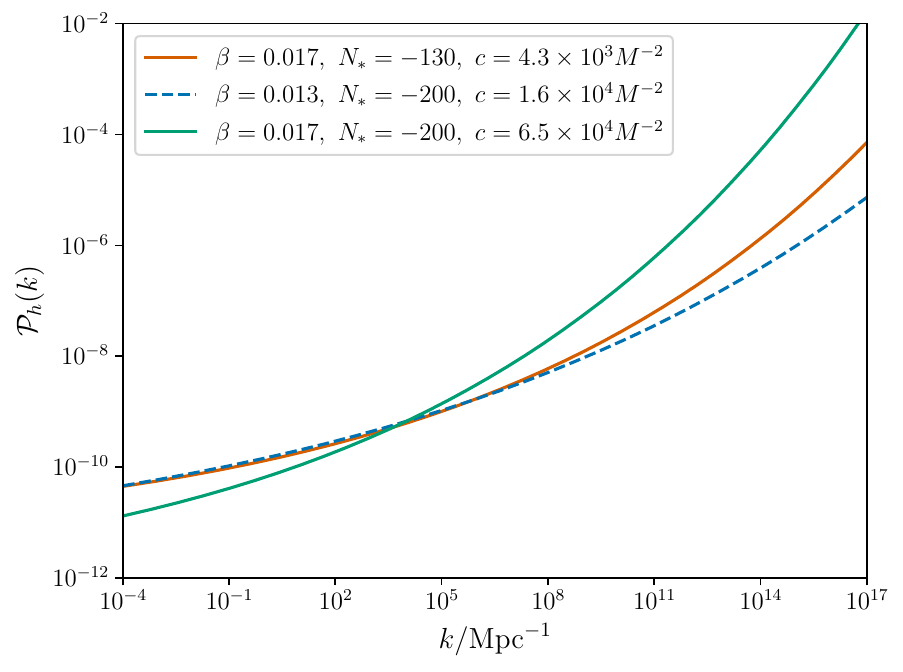}
\caption{ Resulting tensor power spectra, defined as $\mathcal{P}_h(k)=k^3/(2\pi^2)\left(|h_L(k)|^2+|h_R(k)|^2\right)$, for three representative parameter sets. Across the full range of scales shown, the spectra are dominated by the contribution from the left-handed polarization state. }
\label{fig2}
\end{figure}

Figure 1 compares the predicted scalar spectral index $n_s$ and tensor-to-scalar ratio $r$ with the observational constraints. For a fixed value of $\beta$, decreasing $N_\ast$ to more negative values suppresses $r$ and shifts $n_s$ toward its asymptotic value $1-2\beta$. Consequently, parameter choices with $N_\ast$ close to $-60$ generally predict a relatively large tensor-to-scalar ratio and are mostly disfavored, whereas more negative values of $N_\ast$ provide better agreement with the observations. The Planck+BICEP/Keck constraints favor comparatively larger values of $\beta$, approximately $\beta\sim0.013$-$0.02$, while the data combination including ACT favors a larger scalar spectral index and hence somewhat smaller values, approximately $\beta\sim0.009$-$0.016$. Overall, the constant-roll inflation admits parameter regions compatible with both observational data sets, although their preferred ranges of $\beta$ differ.

\section{Broadband Chiral Primordial Gravitational Waves}

In this section, we embed the constant-roll scenario developed above into a parity-violating extension of STG and investigate its implications for primordial GWs.

\subsection{Tensor perturbations in parity-violating symmetric teleparallel gravity}

In STG, the gravitational interaction is encoded in the nonvanishing nonmetricity tensor,
\begin{align}
    Q_{\alpha\mu\nu}\equiv\nabla_\alpha g_{\mu\nu},
\end{align}
while both the curvature and torsion of the affine connection vanish.
STEGR is formulated in terms of the nonmetricity scalar $\mathbb{Q}$, and its action differs from the Einstein--Hilbert action only by a boundary term, thereby yielding dynamically equivalent gravitational field equations. A parity-violating extension of STEGR coupled to a canonical scalar field can be written schematically as \cite{Li:2022vtn}
\begin{align}
S=\int {\rm d}^4x\sqrt{-g}&\left[\frac{M^2_{\rm Pl}}{2}\mathbb{Q}+\sum_{i=1}^7 c_i(\phi,\nabla^\mu \phi \nabla_\mu \phi) \mathcal{M}_i \right. \nonumber  \\ 
& \qquad \left. -\frac{1}{2} g^{\mu\nu}\partial_\mu\phi\partial_\nu\phi-V(\phi)\right].
\end{align}
where $\mathcal{M}_i$ denote the seven independent parity-odd operators constructed from the nonmetricity tensor and the first derivatives of the scalar field. The coupling functions $c_i$ characterize the strength of the corresponding parity-violating interactions. 

\begin{figure*}
\centering
\includegraphics[width=0.95\textwidth]{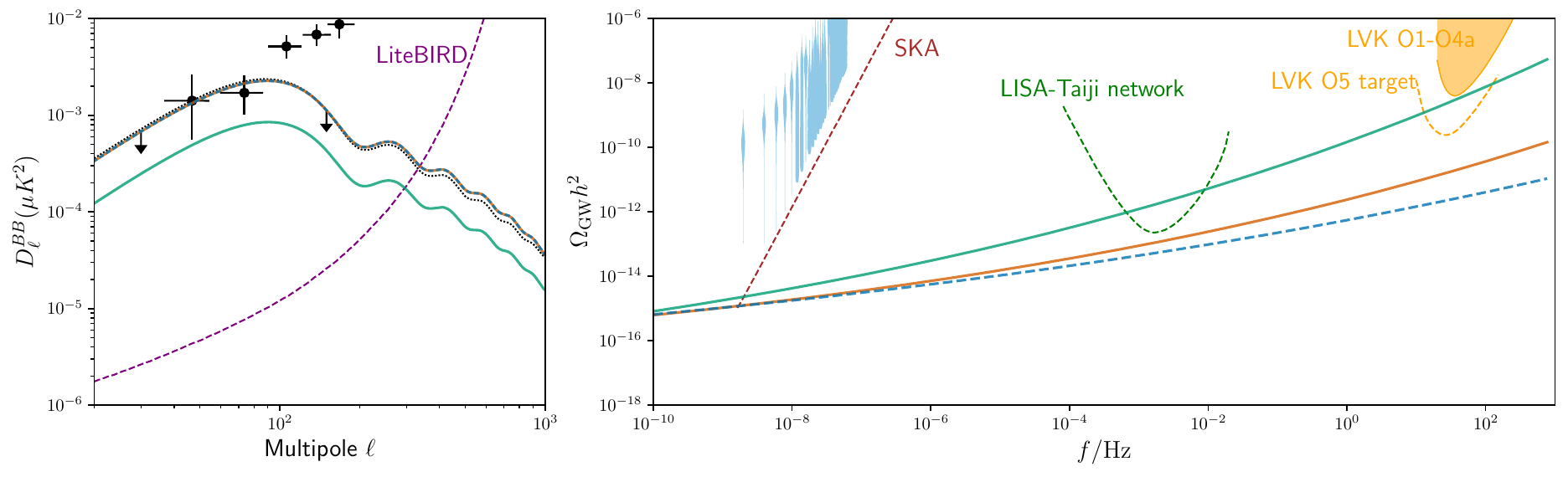}
\caption{{\it Left panel}: Predicted CMB B-mode angular power spectra, $D^{BB}_\ell = \ell (\ell +1 )C^{BB}_\ell/ (2\pi)$. The black data points show the CMB B-mode bandpowers measured by BICEP/Keck 2018~\cite{BICEP:2021xfz}. The purple dashed curve indicates the forecast sensitivity of the future CMB satellite LiteBIRD~\cite{Ishino:2016izb}. The black dotted curve corresponds to a scale-invariant primordial tensor spectrum with $r=0.036$. The remaining curves show the BB spectra generated by the primordial tensor power spectra presented in Fig. \ref{fig2}.  {\it Right panel}: Predicted present-day energy density spectra of primordial GWs. The light-blue violin plots show the free-spectrum posterior distributions inferred from the NANOGrav 15-year data set~\cite{NANOGrav:2023hvm}.  The brown dashed line denotes the projected sensitivity of the Square Kilometre Array (SKA)~\cite{Janssen:2014dka}, while the green dashed curve represents the power-law integrated sensitivity curve of the LISA--Taiji network~\cite{Chen:2024ikn}. The orange shaded region shows the current constraints from the LVK O1--O4a stochastic background search, and the orange dashed curve denotes the target $2\sigma$ power-law integrated sensitivity curve for the O5 observing run~\cite{LIGOScientific:2025bgj}. The remaining curves correspond to the current GW spectra generated by the primordial tensor power spectra shown in Fig.~\ref{fig2}.}
\label{fig3}
\end{figure*}

On a spatially flat FRW background, the parity-violating operators modify neither the homogeneous background evolution nor the linear scalar perturbations. Consequently, the constant-roll background solution and the scalar power spectrum derived in the preceding section remain unchanged. The parity-violating effects instead manifest themselves in the vector and tensor sectors. In this work, we restrict our analysis to linear tensor perturbations. We introduce transverse and traceless tensor perturbations $h_{ij}$ through
\begin{align}
    {\rm d}s^2=-{\rm d}t^2
    +a^2(t)\left(\delta_{ij}+h_{ij}\right)
    {\rm d}x^i{\rm d}x^j.
\end{align}
The tensor perturbations can be decomposed into circular polarization modes as
\begin{align}
    h_{ij}(t,\bm{x})
    =\sum_{A=L,R}\int\frac{{\rm d}^3\bm{k}}{(2\pi)^{3/2}}
    h_A(t,\bm{k})e^A_{ij}(\bm{k})
    e^{i\bm{k}\cdot\bm{x}},
\end{align}
where $A=L,R$ labels the left- and right-handed polarization states, with $\lambda_L=-1$ and $\lambda_R=+1$. Restricting our attention to constant coupling coefficients, the mode function $h_A(t,\bm{k})$ then obeys
\begin{align}
    \ddot h_A+3H\dot h_A
    +\frac{k}{a}\left[
        \frac{k}{a}
        +\lambda_A\frac{c}{M_{\rm Pl}^2}
        \left(H\dot\phi^2+\dot\phi\ddot\phi\right)
    \right]h_A=0,
\end{align}
with $ c = 8(2c_1-c_5)$. Using the constant-roll condition, $\ddot\phi=\beta H\dot\phi$, this equation becomes
\begin{align}\label{EoM_tensor}
    \ddot h_A+3H\dot h_A
    +\frac{k}{a}\left[
        \frac{k}{a}
        +\lambda_A\frac{c(1+\beta)}{M_{\rm Pl}^2}
        H\dot\phi^2
    \right]h_A=0.
\end{align}
Accordingly, the two circular polarization modes obey helicity-dependent dispersion relations,
\begin{align}
    \omega_A^2(t,k)
    =\frac{k}{a}\left[
        \frac{k}{a}+\lambda_A\Lambda(t)
    \right],
    \qquad
    \Lambda(t)\equiv
    \frac{c(1+\beta)}{M_{\rm Pl}^2}H\dot\phi^2.
\end{align}
The dependence on $\lambda_A$ implies that the two polarization states propagate with different velocities, giving rise to velocity birefringence. Furthermore, for the polarization satisfying $\lambda_A\Lambda<0$, the effective frequency squared becomes negative when $k/a<|\Lambda(t)|$. Without loss of generality, we take $c>0$. In this case, the left-handed polarization state can develop a negative effective frequency squared, $\omega_L^2<0$. In particular, if $|\Lambda|\gtrsim H$, the corresponding mode functions undergo a tachyonic instability and are exponentially amplified. 
At CMB scales, if the parity-violating correction is non-negligible, the standard expression for $r$ derived in the preceding section is no longer applicable, and the tensor power spectrum must instead be computed from the modified evolution equations for the two polarization states.

\begin{figure}
\includegraphics[width=1\columnwidth]{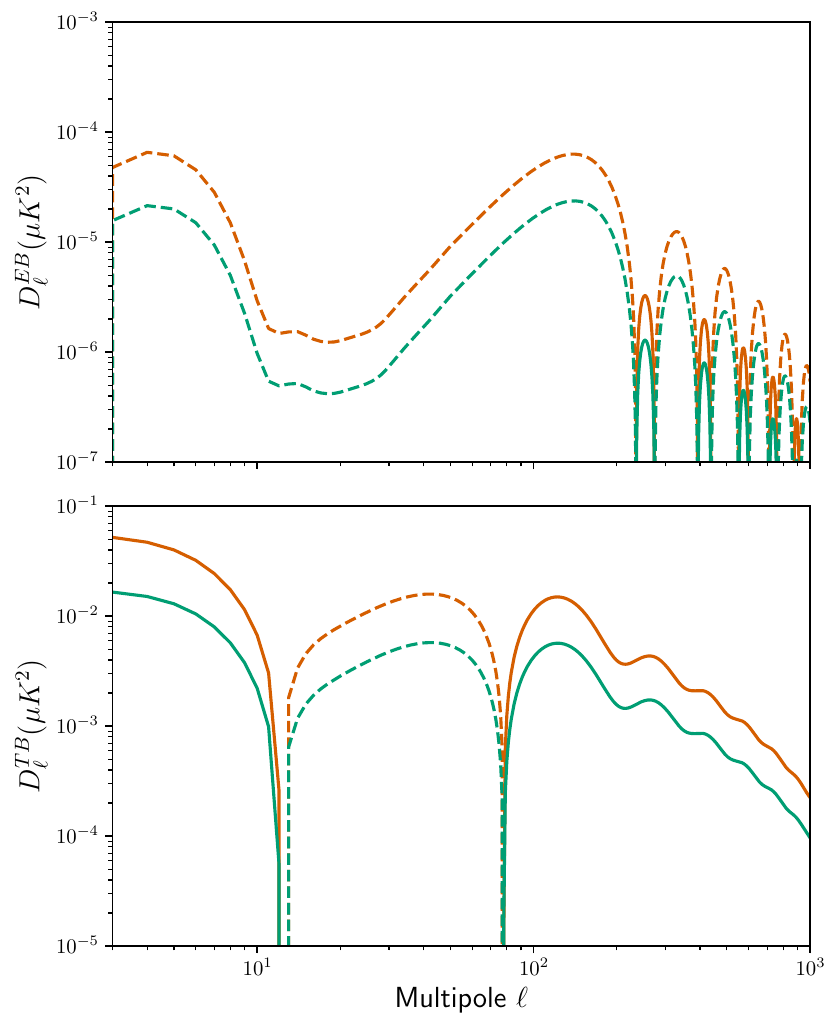}
\caption{ Predicted CMB cross-correlation spectra, $D^{XY}_\ell = \ell (\ell +1 )C^{XY}_\ell/ (2\pi)$ with $XY=\{TB,EB\}$. The absolute values of the spectra are shown, with dashed segments indicating regions where the corresponding spectra are negative. The orange-red curves correspond to $(\beta=0.017, N_\ast=-130, c=4.3\times10^3M^{-2})$; the parameter set $(\beta=0.013, N_\ast=-200, c=1.6\times10^4M^{-2})$ yields similar results. The blue-green curves correspond to $(\beta=0.017, N_\ast=-200, c=6.5\times10^4M^{-2})$.}
\label{fig4}
\end{figure}

\subsection{CMB and multi-band GW phenomenology}
In this subsection, we numerically solve Eqs.~\eqref{background_equation} and \eqref{EoM_tensor} simultaneously to investigate the phenomenology of primordial GWs on both CMB and sub-CMB scales.

In Fig. \ref{fig2}, we present the resulting tensor power spectra,
\begin{align}
    \mathcal{P}_h(k) = \mathcal{P}^L_h(k) + \mathcal{P}^R_h(k) = \frac{k^3}{2\pi^2}\left(|h_L(k)|^2+|h_R(k)|^2\right),
\end{align}
for three representative parameter sets, such as $(\beta=0.017, N_\ast=-130, c=4.3\times10^3M^{-2})$ with $M=5.8\times10^{-6}M_{\rm Pl}$, $(\beta=0.013, N_\ast=-200, c=1.6\times10^4M^{-2})$ with $M=3.4\times10^{-6}M_{\rm Pl}$, and $(\beta=0.017, N_\ast=-200, c=6.5\times10^4M^{-2})$ with $M=1.7\times10^{-6}M_{\rm Pl}$. For all three parameter sets, the tensor power spectrum is dominated by the left-handed polarization, i.e., $\mathcal{P}_h(k) \simeq \mathcal{P}^L_h(k)$, over essentially the entire range of scales shown, extending from CMB to small scales. This indicates that the parity-violating interactions are already significant on CMB scales and generates a highly chiral primordial tensor spectrum. As the wavenumber increases, the corresponding modes encounter the instability at later stages of inflation, when the inflaton velocity is larger. The tachyonic amplification of the left-handed mode therefore becomes progressively more efficient, producing a pronounced blue tilt and a rapid enhancement of the tensor power toward smaller scales. By contrast, the right-handed contribution remains strongly suppressed and is negligible in the total tensor spectrum. This scale-dependent amplification applies only to modes that encounter the instability before the constant-roll phase terminates at $\phi_0$, and therefore the exit introduces a physical ultraviolet cutoff in the enhanced spectrum.

For each parameter set, we choose the coupling $c$ to be the largest value consistent with both large- and small-scale GW constraints. On large scales, primordial tensor perturbations source the CMB B-mode angular power spectrum~\cite{Saito:2007kt,Gluscevic:2010vv},
\begin{align}
    C^{BB}_\ell = 4\pi \int {\rm d}(\ln k) \left[\mathcal{P}^L_h(k)+\mathcal{P}^R_h(k)\right] \left[\Delta_\ell^B(k)\right]^2,
\end{align}
where $\Delta_\ell^B(k)$ denotes the B-mode radiation transfer function. The resulting spectra are compared with the CMB B-mode bandpowers measured by BICEP/Keck 2018~\cite{BICEP:2021xfz} in the left panel of Fig.~\ref{fig3}. On small scales, the corresponding present-day GW energy density spectrum is approximately given by~\cite{Guzzetti:2016mkm,Inomata:2021zel}
\begin{align}
    \Omega_{\rm GW}(f)h^2 \simeq 6.8\times10^{-7}\mathcal{P}_h(k),
\end{align}
where the observed frequency $f$ is related to the comoving wave number $k$ through $f=1.546\times10^{-15}(k/{\rm Mpc}^{-1})\rm Hz$. The high-frequency portion of the resulting spectra, lying within the LIGO-Virgo-KAGRA (LVK) frequency band, is constrained by the stochastic background search in LVK O1-O4a runs~\cite{LIGOScientific:2025bgj}, as shown in the right panel.

The parameters $\beta$ and $N_\ast$ determine the scale dependence of the primordial tensor spectrum. As shown in Fig.~\ref{fig2}, a larger $\beta$ or a more negative $N_\ast$ generally produces a steeper blue tilt. Consequently, the upper bound on $c$ for the first two parameter sets is primarily determined by the CMB B-mode measurements, whereas that for the third set is mainly imposed by the LVK O1-O4a constraint, as illustrated in Fig.~\ref{fig3}. The CMB and interferometer observations therefore provide complementary probes: the former constrain the tensor amplitude on large scales, while the latter restrict its blue enhancement on small scales. On CMB scales, the B-mode signals predicted for all three parameter sets lie within the projected sensitivity of future experiments such as LiteBIRD. By contrast, the small-scale GW signals for the first two parameter sets remain below the projected sensitivities of SKA, the LISA--Taiji network, and the LVK O5 run. For the third parameter set, however, the signal may be accessible to both the LISA--Taiji network and LVK O5. Moreover, the LISA-Taiji network could probe the chirality of the predicted GW signal, which is nearly fully one-handed.
In summary, future multiband GW observations could provide a powerful probe of the parity-violating interactions and constant-roll inflationary dynamics across the phenomenologically viable parameter space of this model.

Furthermore, a chiral primordial GW background generates non-vanishing cross-correlation spectra between the CMB temperature and B-mode (TB), as well as between E-mode and B-mode (EB), which are given by~\cite{Saito:2007kt,Gluscevic:2010vv}
\begin{align}
    C^{XB}_\ell = 4\pi \int {\rm d}(\ln k) \left[\mathcal{P}^L_h(k)-\mathcal{P}^R_h(k)\right] \Delta_\ell^X(k)\Delta_\ell^B(k),
\end{align}
where $X=\{T,E\}$, and $\Delta_\ell^{T/E}(k)$ are the corresponding radiation transfer functions. As illustrated in Fig.~\ref{fig4}, the nearly one-handed tensor spectra predicted by our benchmark parameter sets generate nonvanishing $TB$ and $EB$ spectra. As an illustrative reference, the Fisher forecast of Ref.~\cite{Gluscevic:2010vv} found that a CMBPol-like experiment~\cite{CMBPolStudyTeam:2008rgp} could achieve marginal sensitivity to a maximally chiral tensor background for a tensor amplitude of order $r\sim10^{-2}$, with most of the constraining power arising from the large-angle $TB$ spectrum. Although this estimate cannot be directly applied to the scale-dependent tensor spectrum considered here, it suggests that the near-maximal chirality, together with the non-negligible tensor power predicted on CMB scales, could make the $TB$ and $EB$ correlations a complementary observational channel for testing the model with future CMB polarization measurements.

\section{Conclusions}
We have investigated primordial GWs generated when constant-roll inflation, characterized by $\ddot\phi=\beta H \dot\phi$, is embedded in a parity-violating extension of STEGR. Since the parity-violating terms affect neither the homogeneous background nor the linear scalar perturbations, the scalar-sector predictions remain identical to those of the conventional canonical constant-roll model. For the exact potential $V(\phi)$ considered here, inflation must terminate before the field reaches the critical value $\phi_{\rm c}$, at which $V(\phi_{\rm c})=0$. We parameterize this exit by introducing a cutoff $\phi_0<\phi_{\rm c}$ and define the {\it e}-folding number $N$ relative to $\phi_{\rm c}$, with $N_\ast$ denoting its value when the CMB pivot scale exits the Hubble horizon. The Planck+BICEP/Keck data favor approximately $\beta\simeq0.013$--$0.02$, whereas the data combination including ACT prefers the slightly smaller range $\beta\simeq0.009$--$0.016$. The constant-roll inflation therefore contains observationally viable regions of parameter space for both data combinations.

The tensor sector differs qualitatively from that of conventional canonical constant-roll inflation. For constant parity-violating couplings, the dispersion relation of each circular polarization contains the helicity-dependent scale, $\Lambda=c(1+\beta)H\dot\phi^2/M_{\rm Pl}^2$. Taking $c>0$, the effective frequency squared of the left-handed mode becomes negative when $k/a<\Lambda$. If $\Lambda$ is comparable to or larger than the Hubble parameter, the mode experiences a tachyonic instability. As the inflaton velocity increases, modes with larger wavenumbers encounter an increasingly efficient instability at later stages of inflation. Consequently, our numerical results exhibit a strongly blue tensor spectrum that is nearly entirely dominated by the left-handed polarization state across the full range from CMB to interferometer scales. This broadband enhancement distinguishes the present scenario from models in which a localized feature in the inflationary potential amplifies only a restricted range of modes.

For each benchmark parameter set, we choose the coupling parameter $c$ close to the largest value consistent with the current CMB $B$-mode measurements and the LVK O1--O4a constraint on the stochastic GW background. These constraints highlight the complementarity between large- and small-scale observations. Within the parameter range considered here, a larger $\beta$ or a more negative $N_\ast$ produces a steeper blue tilt and consequently strengthens the small-scale constraints. The predicted $B$-mode spectra fall within the projected sensitivity of LiteBIRD, while the high-frequency signal of the most strongly enhanced benchmark may be accessible to both the LISA--Taiji network and LVK O5. The strong chirality of primordial GWs also generates nonvanishing $TB$ and $EB$ correlations, which could provide a complementary observational channel for future CMB polarization measurements. Taken together, coordinated observations across CMB and interferometer scales could provide a powerful test of parity-violating gravity during inflation.

\begin{acknowledgments}
This work is supported by the National Natural Science Foundation of China Grants No. 12305057 and No. 12405074.
\end{acknowledgments}

\bibliography{references}  

@article{Starobinsky:1980te,
    author = "Starobinsky, Alexei A.",
    editor = "Khalatnikov, I. M. and Mineev, V. P.",
    title = "{A New Type of Isotropic Cosmological Models Without Singularity}",
    doi = "10.1016/0370-2693(80)90670-X",
    journal = "Phys. Lett. B",
    volume = "91",
    pages = "99--102",
    year = "1980"
}

@article{Guth:1980zm,
    author = "Guth, Alan H.",
    editor = "Fang, Li-Zhi and Ruffini, R.",
    title = "{The Inflationary Universe: A Possible Solution to the Horizon and Flatness Problems}",
    reportNumber = "SLAC-PUB-2576",
    doi = "10.1103/PhysRevD.23.347",
    journal = "Phys. Rev. D",
    volume = "23",
    pages = "347--356",
    year = "1981"
}

@article{Linde:1981mu,
    author = "Linde, Andrei D.",
    editor = "Fang, Li-Zhi and Ruffini, R.",
    title = "{A New Inflationary Universe Scenario: A Possible Solution of the Horizon, Flatness, Homogeneity, Isotropy and Primordial Monopole Problems}",
    reportNumber = "LEBEDEV-81-229",
    doi = "10.1016/0370-2693(82)91219-9",
    journal = "Phys. Lett. B",
    volume = "108",
    pages = "389--393",
    year = "1982"
}

@article{Albrecht:1982wi,
    author = "Albrecht, Andreas and Steinhardt, Paul J.",
    editor = "Fang, Li-Zhi and Ruffini, R.",
    title = "{Cosmology for Grand Unified Theories with Radiatively Induced Symmetry Breaking}",
    reportNumber = "UPR-0185T",
    doi = "10.1103/PhysRevLett.48.1220",
    journal = "Phys. Rev. Lett.",
    volume = "48",
    pages = "1220--1223",
    year = "1982"
}

@article{Linde:1983gd,
    author = "Linde, Andrei D.",
    title = "{Chaotic Inflation}",
    doi = "10.1016/0370-2693(83)90837-7",
    journal = "Phys. Lett. B",
    volume = "129",
    pages = "177--181",
    year = "1983"
}

@article{Guth:1982ec,
    author = "Guth, Alan H. and Pi, S. Y.",
    title = "{Fluctuations in the New Inflationary Universe}",
    doi = "10.1103/PhysRevLett.49.1110",
    journal = "Phys. Rev. Lett.",
    volume = "49",
    pages = "1110--1113",
    year = "1982"
}

@article{Starobinsky:1982ee,
    author = "Starobinsky, Alexei A.",
    title = "{Dynamics of Phase Transition in the New Inflationary Universe Scenario and Generation of Perturbations}",
    doi = "10.1016/0370-2693(82)90541-X",
    journal = "Phys. Lett. B",
    volume = "117",
    pages = "175--178",
    year = "1982"
}

@article{Bardeen:1983qw,
    author = "Bardeen, James M. and Steinhardt, Paul J. and Turner, Michael S.",
    title = "{Spontaneous Creation of Almost Scale - Free Density Perturbations in an Inflationary Universe}",
    reportNumber = "UPR-0202T, EFI-83-13-CHICAGO",
    doi = "10.1103/PhysRevD.28.679",
    journal = "Phys. Rev. D",
    volume = "28",
    pages = "679",
    year = "1983"
}

@article{Kodama:1984ziu,
    author = "Kodama, Hideo and Sasaki, Misao",
    title = "{Cosmological Perturbation Theory}",
    doi = "10.1143/PTPS.78.1",
    journal = "Prog. Theor. Phys. Suppl.",
    volume = "78",
    pages = "1--166",
    year = "1984"
}

@article{Guzzetti:2016mkm,
    author = "Guzzetti, M. C. and Bartolo, N. and Liguori, M. and Matarrese, S.",
    title = "{Gravitational waves from inflation}",
    eprint = "1605.01615",
    archivePrefix = "arXiv",
    primaryClass = "astro-ph.CO",
    doi = "10.1393/ncr/i2016-10127-1",
    journal = "Riv. Nuovo Cim.",
    volume = "39",
    number = "9",
    pages = "399--495",
    year = "2016"
}

@article{Caprini:2018mtu,
    author = "Caprini, Chiara and Figueroa, Daniel G.",
    title = "{Cosmological backgrounds of gravitational waves}",
    eprint = "1801.04268",
    archivePrefix = "arXiv",
    primaryClass = "astro-ph.CO",
    doi = "10.1088/1361-6382/aac608",
    journal = "Class. Quant. Grav.",
    volume = "35",
    number = "16",
    pages = "163001",
    year = "2018"
}

@article{Campeti:2020xwn,
    author = "Campeti, Paolo and Komatsu, Eiichiro and Poletti, Davide and Baccigalupi, Carlo",
    title = "{Measuring the spectrum of primordial gravitational waves with CMB, PTA and Laser Interferometers}",
    eprint = "2007.04241",
    archivePrefix = "arXiv",
    primaryClass = "astro-ph.CO",
    doi = "10.1088/1475-7516/2021/01/012",
    journal = "JCAP",
    volume = "01",
    pages = "012",
    year = "2021"
}

@article{BICEP:2021xfz,
    author = "Ade, P. A. R. and others",
    collaboration = "BICEP, Keck",
    title = "{Improved Constraints on Primordial Gravitational Waves using Planck, WMAP, and BICEP/Keck Observations through the 2018 Observing Season}",
    eprint = "2110.00483",
    archivePrefix = "arXiv",
    primaryClass = "astro-ph.CO",
    doi = "10.1103/PhysRevLett.127.151301",
    journal = "Phys. Rev. Lett.",
    volume = "127",
    number = "15",
    pages = "151301",
    year = "2021"
}

@article{AtacamaCosmologyTelescope:2025nti,
    author = "Calabrese, Erminia and others",
    collaboration = "Atacama Cosmology Telescope",
    title = "{The Atacama Cosmology Telescope: DR6 constraints on extended cosmological models}",
    eprint = "2503.14454",
    archivePrefix = "arXiv",
    primaryClass = "astro-ph.CO",
    reportNumber = "FERMILAB-PUB-25-0157-PPD",
    doi = "10.1088/1475-7516/2025/11/063",
    journal = "JCAP",
    volume = "11",
    pages = "063",
    year = "2025"
}

@article{Cook:2011hg,
    author = "Cook, Jessica L. and Sorbo, Lorenzo",
    title = "{Particle production during inflation and gravitational waves detectable by ground-based interferometers}",
    eprint = "1109.0022",
    archivePrefix = "arXiv",
    primaryClass = "astro-ph.CO",
    doi = "10.1103/PhysRevD.85.023534",
    journal = "Phys. Rev. D",
    volume = "85",
    pages = "023534",
    year = "2012",
    note = "[Erratum: Phys.Rev.D 86, 069901 (2012)]"
}

@article{Mukohyama:2014gba,
    author = "Mukohyama, Shinji and Namba, Ryo and Peloso, Marco and Shiu, Gary",
    title = "{Blue Tensor Spectrum from Particle Production during Inflation}",
    eprint = "1405.0346",
    archivePrefix = "arXiv",
    primaryClass = "astro-ph.CO",
    doi = "10.1088/1475-7516/2014/08/036",
    journal = "JCAP",
    volume = "08",
    pages = "036",
    year = "2014"
}

@article{Cai:2016ldn,
    author = "Cai, Yong and Wang, Yu-Tong and Piao, Yun-Song",
    title = "{Propagating speed of primordial gravitational waves and inflation}",
    eprint = "1602.05431",
    archivePrefix = "arXiv",
    primaryClass = "astro-ph.CO",
    doi = "10.1103/PhysRevD.94.043002",
    journal = "Phys. Rev. D",
    volume = "94",
    number = "4",
    pages = "043002",
    year = "2016"
}

@article{Bartolo:2016ami,
    author = "Bartolo, Nicola and others",
    title = "{Science with the space-based interferometer LISA. IV: Probing inflation with gravitational waves}",
    eprint = "1610.06481",
    archivePrefix = "arXiv",
    primaryClass = "astro-ph.CO",
    reportNumber = "ACFI-T16-19, UMN-TH-3608-16, CERN-TH-2016-222, KCL-PH-TH-2016-58, IFT-UAM-CSIC-16-104",
    doi = "10.1088/1475-7516/2016/12/026",
    journal = "JCAP",
    volume = "12",
    pages = "026",
    year = "2016"
}

@article{Mylova:2018yap,
    author = {Mylova, Maria and {\"O}zsoy, Ogan and Parameswaran, Susha and Tasinato, Gianmassimo and Zavala, Ivonne},
    title = "{A new mechanism to enhance primordial tensor fluctuations in single field inflation}",
    eprint = "1808.10475",
    archivePrefix = "arXiv",
    primaryClass = "gr-qc",
    doi = "10.1088/1475-7516/2018/12/024",
    journal = "JCAP",
    volume = "12",
    pages = "024",
    year = "2018"
}

@article{Satoh:2007gn,
    author = "Satoh, Masaki and Kanno, Sugumi and Soda, Jiro",
    title = "{Circular Polarization of Primordial Gravitational Waves in String-inspired Inflationary Cosmology}",
    eprint = "0706.3585",
    archivePrefix = "arXiv",
    primaryClass = "astro-ph",
    doi = "10.1103/PhysRevD.77.023526",
    journal = "Phys. Rev. D",
    volume = "77",
    pages = "023526",
    year = "2008"
}

@article{Fu:2020tlw,
    author = "Fu, Chengjie and Liu, Jing and Zhu, Tao and Yu, Hongwei and Wu, Puxun",
    title = "{Resonance instability of primordial gravitational waves during inflation in Chern{\textendash}Simons gravity}",
    eprint = "2006.03771",
    archivePrefix = "arXiv",
    primaryClass = "gr-qc",
    doi = "10.1140/epjc/s10052-021-09001-2",
    journal = "Eur. Phys. J. C",
    volume = "81",
    number = "3",
    pages = "204",
    year = "2021"
}

@article{Cai:2021uup,
    author = "Cai, Rong-Gen and Fu, Chengjie and Yu, Wang-Wei",
    title = "{Parity violation in stochastic gravitational wave background from inflation in Nieh-Yan modified teleparallel gravity}",
    eprint = "2112.04794",
    archivePrefix = "arXiv",
    primaryClass = "astro-ph.CO",
    doi = "10.1103/PhysRevD.105.103520",
    journal = "Phys. Rev. D",
    volume = "105",
    number = "10",
    pages = "103520",
    year = "2022"
}

@article{Fu:2023aab,
    author = "Fu, Chengjie and Liu, Jing and Yang, Xing-Yu and Yu, Wang-Wei and Zhang, Yawen",
    title = "{Explaining pulsar timing array observations with primordial gravitational waves in parity-violating gravity}",
    eprint = "2308.15329",
    archivePrefix = "arXiv",
    primaryClass = "astro-ph.CO",
    doi = "10.1103/PhysRevD.109.063526",
    journal = "Phys. Rev. D",
    volume = "109",
    number = "6",
    pages = "063526",
    year = "2024"
}

@article{Zhai:2025flf,
    author = "Zhai, Rongrong and Fu, Chengjie and Fu, Xiangyun and Wu, Puxun and Yu, Hongwei",
    title = "{Primordial gravitational waves in parity-violating symmetric teleparallel gravity}",
    eprint = "2508.06984",
    archivePrefix = "arXiv",
    primaryClass = "astro-ph.CO",
    doi = "10.1103/hw8n-bsw5",
    journal = "Phys. Rev. D",
    volume = "113",
    number = "6",
    pages = "063517",
    year = "2026"
}

@article{Lue:1998mq,
    author = "Lue, Arthur and Wang, Li-Min and Kamionkowski, Marc",
    title = "{Cosmological signature of new parity violating interactions}",
    eprint = "astro-ph/9812088",
    archivePrefix = "arXiv",
    reportNumber = "CU-TP-926, CAL-675",
    doi = "10.1103/PhysRevLett.83.1506",
    journal = "Phys. Rev. Lett.",
    volume = "83",
    pages = "1506--1509",
    year = "1999"
}

@article{Alexander:2009tp,
    author = "Alexander, Stephon and Yunes, Nicolas",
    title = "{Chern-Simons Modified General Relativity}",
    eprint = "0907.2562",
    archivePrefix = "arXiv",
    primaryClass = "hep-th",
    doi = "10.1016/j.physrep.2009.07.002",
    journal = "Phys. Rept.",
    volume = "480",
    pages = "1--55",
    year = "2009"
}

@article{Wang:2012fi,
    author = "Wang, Anzhong and Wu, Qiang and Zhao, Wen and Zhu, Tao",
    title = "{Polarizing primordial gravitational waves by parity violation}",
    eprint = "1208.5490",
    archivePrefix = "arXiv",
    primaryClass = "astro-ph.CO",
    doi = "10.1103/PhysRevD.87.103512",
    journal = "Phys. Rev. D",
    volume = "87",
    number = "10",
    pages = "103512",
    year = "2013"
}

@article{Alexander:2016hxk,
    author = "Alexander, Stephon H. S.",
    title = "{Inflationary Birefringence and Baryogenesis}",
    eprint = "1604.00703",
    archivePrefix = "arXiv",
    primaryClass = "hep-th",
    doi = "10.1142/S0218271816400137",
    journal = "Int. J. Mod. Phys. D",
    volume = "25",
    number = "11",
    pages = "1640013",
    year = "2016"
}

@article{Bartolo:2017szm,
    author = "Bartolo, Nicola and Orlando, Giorgio",
    title = "{Parity breaking signatures from a Chern-Simons coupling during inflation: the case of non-Gaussian gravitational waves}",
    eprint = "1706.04627",
    archivePrefix = "arXiv",
    primaryClass = "astro-ph.CO",
    doi = "10.1088/1475-7516/2017/07/034",
    journal = "JCAP",
    volume = "07",
    pages = "034",
    year = "2017"
}

@article{Qiao:2019hkz,
    author = "Qiao, Jin and Zhu, Tao and Zhao, Wen and Wang, Anzhong",
    title = "{Polarized primordial gravitational waves in the ghost-free parity-violating gravity}",
    eprint = "1911.01580",
    archivePrefix = "arXiv",
    primaryClass = "astro-ph.CO",
    doi = "10.1103/PhysRevD.101.043528",
    journal = "Phys. Rev. D",
    volume = "101",
    number = "4",
    pages = "043528",
    year = "2020"
}

@article{Zhu:2022dfq,
    author = "Zhu, Tao and Zhao, Wen and Wang, Anzhong",
    title = "{Polarized primordial gravitational waves in spatial covariant gravities}",
    eprint = "2210.05259",
    archivePrefix = "arXiv",
    primaryClass = "gr-qc",
    doi = "10.1103/PhysRevD.107.024031",
    journal = "Phys. Rev. D",
    volume = "107",
    number = "2",
    pages = "024031",
    year = "2023"
}

@article{Saito:2007kt,
    author = "Saito, Shun and Ichiki, Kiyotomo and Taruya, Atsushi",
    title = "{Probing polarization states of primordial gravitational waves with CMB anisotropies}",
    eprint = "0705.3701",
    archivePrefix = "arXiv",
    primaryClass = "astro-ph",
    doi = "10.1088/1475-7516/2007/09/002",
    journal = "JCAP",
    volume = "09",
    pages = "002",
    year = "2007"
}

@article{Gluscevic:2010vv,
    author = "Gluscevic, Vera and Kamionkowski, Marc",
    title = "{Testing Parity-Violating Mechanisms with Cosmic Microwave Background Experiments}",
    eprint = "1002.1308",
    archivePrefix = "arXiv",
    primaryClass = "astro-ph.CO",
    doi = "10.1103/PhysRevD.81.123529",
    journal = "Phys. Rev. D",
    volume = "81",
    pages = "123529",
    year = "2010"
}

@article{Seto:2020zxw,
    author = "Seto, Naoki",
    title = "{Measuring Parity Asymmetry of Gravitational Wave Backgrounds with a Heliocentric Detector Network in the mHz Band}",
    eprint = "2009.02928",
    archivePrefix = "arXiv",
    primaryClass = "gr-qc",
    doi = "10.1103/PhysRevLett.125.251101",
    journal = "Phys. Rev. Lett.",
    volume = "125",
    pages = "251101",
    year = "2020"
}

@article{Orlando:2020oko,
    author = "Orlando, Giorgio and Pieroni, Mauro and Ricciardone, Angelo",
    title = "{Measuring Parity Violation in the Stochastic Gravitational Wave Background with the LISA-Taiji network}",
    eprint = "2011.07059",
    archivePrefix = "arXiv",
    primaryClass = "astro-ph.CO",
    doi = "10.1088/1475-7516/2021/03/069",
    journal = "JCAP",
    volume = "03",
    pages = "069",
    year = "2021"
}

@article{Chen:2024ikn,
    author = "Chen, Ju and Liu, Chang and Zhang, Yun-Long",
    title = "{Circularly polarized gravitational wave background search with a network of space-borne triangular detectors}",
    eprint = "2410.18916",
    archivePrefix = "arXiv",
    primaryClass = "gr-qc",
    doi = "10.1088/1475-7516/2025/05/050",
    journal = "JCAP",
    volume = "05",
    pages = "050",
    year = "2025"
}

@article{Nester:1998mp,
    author = "Nester, James M. and Yo, Hwei-Jang",
    title = "{Symmetric teleparallel general relativity}",
    eprint = "gr-qc/9809049",
    archivePrefix = "arXiv",
    reportNumber = "NCU-CCS-980904",
    journal = "Chin. J. Phys.",
    volume = "37",
    pages = "113",
    year = "1999"
}

@article{BeltranJimenez:2019esp,
    author = "Beltr{\'a}n Jim{\'e}nez, Jose and Heisenberg, Lavinia and Koivisto, Tomi S.",
    title = "{The Geometrical Trinity of Gravity}",
    eprint = "1903.06830",
    archivePrefix = "arXiv",
    primaryClass = "hep-th",
    doi = "10.3390/universe5070173",
    journal = "Universe",
    volume = "5",
    number = "7",
    pages = "173",
    year = "2019"
}

@article{Capozziello:2022zzh,
    author = "Capozziello, Salvatore and De Falco, Vittorio and Ferrara, Carmen",
    title = "{Comparing equivalent gravities: common features and differences}",
    eprint = "2208.03011",
    archivePrefix = "arXiv",
    primaryClass = "gr-qc",
    doi = "10.1140/epjc/s10052-022-10823-x",
    journal = "Eur. Phys. J. C",
    volume = "82",
    number = "10",
    pages = "865",
    year = "2022"
}

@article{Li:2021mdp,
    author = "Li, Mingzhe and Zhao, Dehao",
    title = "{A simple parity violating model in the symmetric teleparallel gravity and its cosmological perturbations}",
    eprint = "2108.01337",
    archivePrefix = "arXiv",
    primaryClass = "gr-qc",
    reportNumber = "USTC-ICTS/PCFT-21-31",
    doi = "10.1016/j.physletb.2022.136968",
    journal = "Phys. Lett. B",
    volume = "827",
    pages = "136968",
    year = "2022"
}

@article{Li:2022vtn,
    author = "Li, Mingzhe and Tong, Yeheng and Zhao, Dehao",
    title = "{Possible consistent model of parity violations in the symmetric teleparallel gravity}",
    eprint = "2203.06912",
    archivePrefix = "arXiv",
    primaryClass = "gr-qc",
    doi = "10.1103/PhysRevD.105.104002",
    journal = "Phys. Rev. D",
    volume = "105",
    number = "10",
    pages = "104002",
    year = "2022"
}

@article{Motohashi:2014ppa,
    author = "Motohashi, Hayato and Starobinsky, Alexei A. and Yokoyama, Jun'ichi",
    title = "{Inflation with a constant rate of roll}",
    eprint = "1411.5021",
    archivePrefix = "arXiv",
    primaryClass = "astro-ph.CO",
    reportNumber = "RESCEU-51-14",
    doi = "10.1088/1475-7516/2015/09/018",
    journal = "JCAP",
    volume = "09",
    pages = "018",
    year = "2015"
}

@article{Motohashi:2017aob,
    author = "Motohashi, Hayato and Starobinsky, Alexei A.",
    title = "{Constant-roll inflation: confrontation with recent observational data}",
    eprint = "1702.05847",
    archivePrefix = "arXiv",
    primaryClass = "astro-ph.CO",
    doi = "10.1209/0295-5075/117/39001",
    journal = "EPL",
    volume = "117",
    number = "3",
    pages = "39001",
    year = "2017"
}

@article{Cicciarella:2017nls,
    author = "Cicciarella, Francesco and Mabillard, Joel and Pieroni, Mauro",
    title = "{New perspectives on constant-roll inflation}",
    eprint = "1709.03527",
    archivePrefix = "arXiv",
    primaryClass = "astro-ph.CO",
    reportNumber = "IFT-UAM:CSIC-17-083",
    doi = "10.1088/1475-7516/2018/01/024",
    journal = "JCAP",
    volume = "01",
    pages = "024",
    year = "2018"
}

@article{Yi:2017mxs,
    author = "Yi, Zhu and Gong, Yungui",
    title = "{On the constant-roll inflation}",
    eprint = "1712.07478",
    archivePrefix = "arXiv",
    primaryClass = "gr-qc",
    doi = "10.1088/1475-7516/2018/03/052",
    journal = "JCAP",
    volume = "03",
    pages = "052",
    year = "2018"
}

@article{Gao:2018cpp,
    author = "Gao, Qing",
    title = "{The observational constraint on constant-roll inflation}",
    eprint = "1802.01986",
    archivePrefix = "arXiv",
    primaryClass = "gr-qc",
    doi = "10.1007/s11433-018-9197-2",
    journal = "Sci. China Phys. Mech. Astron.",
    volume = "61",
    number = "7",
    pages = "070411",
    year = "2018"
}

@article{GalvezGhersi:2018haa,
    author = "Galvez Ghersi, Jose T. and Zucca, Alex and Frolov, Andrei V.",
    title = "{Observational Constraints on Constant Roll Inflation}",
    eprint = "1808.01325",
    archivePrefix = "arXiv",
    primaryClass = "astro-ph.CO",
    reportNumber = "SCG-2018-15",
    doi = "10.1088/1475-7516/2019/05/030",
    journal = "JCAP",
    volume = "05",
    pages = "030",
    year = "2019"
}

@article{Gao:2019sbz,
    author = "Gao, Qing and Gong, Yungui and Yi, Zhu",
    title = "{On the constant-roll inflation with large and small $\eta_H$}",
    eprint = "1901.04646",
    archivePrefix = "arXiv",
    primaryClass = "gr-qc",
    doi = "10.3390/universe5110215",
    journal = "Universe",
    volume = "5",
    number = "11",
    pages = "215",
    year = "2019"
}

@article{Lin:2019fcz,
    author = "Lin, Wei-Chen and Morse, Michael J. P. and Kinney, William H.",
    title = "{Dynamical Analysis of Attractor Behavior in Constant Roll Inflation}",
    eprint = "1904.06289",
    archivePrefix = "arXiv",
    primaryClass = "astro-ph.CO",
    doi = "10.1088/1475-7516/2019/09/063",
    journal = "JCAP",
    volume = "09",
    pages = "063",
    year = "2019"
}

@article{Motohashi:2019rhu,
    author = "Motohashi, Hayato and Mukohyama, Shinji and Oliosi, Michele",
    title = "{Constant Roll and Primordial Black Holes}",
    eprint = "1910.13235",
    archivePrefix = "arXiv",
    primaryClass = "gr-qc",
    reportNumber = "YITP-19-92, IPMU19-0139",
    doi = "10.1088/1475-7516/2020/03/002",
    journal = "JCAP",
    volume = "03",
    pages = "002",
    year = "2020"
}

@inproceedings{Motohashi:2025qgd,
    author = "Motohashi, Hayato",
    title = "{Constant-Roll Inflation}",
    eprint = "2504.16757",
    archivePrefix = "arXiv",
    primaryClass = "astro-ph.CO",
    month = "4",
    year = "2025"
}

@article{Ishino:2016izb,
    author = "Ishino, H. and others",
    editor = "MacEwen, Howard A. and Fazio, Giovanni G. and Lystrup, Makenzie and Batalha, Natalie and Siegler, Nicholas and Tong, Edward C.",
    title = "{LiteBIRD: lite satellite for the study of B-mode polarization and inflation from cosmic microwave background radiation detection}",
    doi = "10.1117/12.2231995",
    journal = "Proc. SPIE Int. Soc. Opt. Eng.",
    volume = "9904",
    pages = "99040X",
    year = "2016"
}

@article{NANOGrav:2023hvm,
    author = "Afzal, Adeela and others",
    collaboration = "NANOGrav",
    title = "{The NANOGrav 15 yr Data Set: Search for Signals from New Physics}",
    eprint = "2306.16219",
    archivePrefix = "arXiv",
    primaryClass = "astro-ph.HE",
    reportNumber = "FERMILAB-PUB-23-589-T",
    doi = "10.3847/2041-8213/acdc91",
    journal = "Astrophys. J. Lett.",
    volume = "951",
    number = "1",
    pages = "L11",
    year = "2023",
    note = "[Erratum: Astrophys.J.Lett. 971, L27 (2024), Erratum: Astrophys.J. 971, L27 (2024)]"
}

@article{Janssen:2014dka,
    author = "Janssen, Gemma and others",
    editor = "Bourke, Tyler L. and others",
    title = "{Gravitational wave astronomy with the SKA}",
    eprint = "1501.00127",
    archivePrefix = "arXiv",
    primaryClass = "astro-ph.IM",
    doi = "10.22323/1.215.0037",
    journal = "PoS",
    volume = "AASKA14",
    pages = "037",
    year = "2015"
}

@article{LIGOScientific:2025bgj,
    author = "Abac, A. G. and others",
    collaboration = "LIGO Scientific, VIRGO, KAGRA",
    title = "{Upper Limits on the Isotropic Gravitational-Wave Background from the first part of LIGO, Virgo, and KAGRA's fourth Observing Run}",
    eprint = "2508.20721",
    archivePrefix = "arXiv",
    primaryClass = "gr-qc",
    doi = "10.1103/wq57-sjt2",
    journal = "Phys. Rev. D",
    volume = "104",
    number = "12",
    pages = "123525",
    year = "2021"
}

@article{Inomata:2021zel,
    author = "Inomata, Keisuke",
    title = "{Bound on induced gravitational waves during inflation era}",
    eprint = "2109.06192",
    archivePrefix = "arXiv",
    primaryClass = "astro-ph.CO",
    doi = "10.1103/PhysRevD.104.123525",
    journal = "Phys. Rev. D",
    volume = "104",
    number = "12",
    pages = "123525",
    year = "2021"
}

@article{CMBPolStudyTeam:2008rgp,
    author = "Baumann, Daniel and others",
    editor = "Dodelson, Scott and Baumann, Daniel and Cooray, Asantha and Dunkley, Joanna and Fraisse, Aurelien and Jackson, Mark G. and Kogut, Alan and Krauss, Lawrence and Smith, Kendrick and Zaldarriaga, Matias",
    collaboration = "CMBPol Study Team",
    title = "{CMBPol Mission Concept Study: Probing Inflation with CMB Polarization}",
    eprint = "0811.3919",
    archivePrefix = "arXiv",
    primaryClass = "astro-ph",
    reportNumber = "FERMILAB-PUB-08-601-A",
    doi = "10.1063/1.3160885",
    journal = "AIP Conf. Proc.",
    volume = "1141",
    number = "1",
    pages = "10--120",
    year = "2009"
}

\end{document}